\documentclass[12pt]{article}
\usepackage[utf8]{inputenc}
\usepackage[english]{babel}
\usepackage{amssymb,hyperref,graphicx}
\usepackage{epstopdf}

\newcommand{\fnm}{\footnotemark}
\newcommand{\fnt}{\footnotetext}

\begin{document}

\begin{center}

  \large \bf
  Elliptic solutions of generalized Brans-Dicke gravity with a non-universal coupling
\end{center}

\vspace{15pt}

\begin{center}

 \normalsize\bf
        J. M. Alimi\fnm[1]\fnt[1]{jean-michel.alimi@obspm.fr}$^{, a}$,
        A. A. Golubtsova\fnm[2]\fnt[2]{siedhe@gmail.com}$^{, a,b}$
        and   V. Reverdy \fnm[3]\fnt[3]{vincent.reverdy@obspm.fr}$^{, a}$

 \vspace{7pt}

 \it (a) \ \ \ Laboratoire de Univers et Th\'{e}ories (LUTh), Observatoire de Paris,\\
Place Jules Janssen 5, 92190 Meudon, France \\

(b) \ Institute of Gravitation and Cosmology,
 Peoples' Friendship University of Russia,
 6 Miklukho-Maklaya Str.,  Moscow 117198, Russia  \\

 \end{center}
 \vspace{15pt}

\begin{abstract}
We  study a model of the generalized Brans-Dicke gravity  presented in both the Jordan and
in the Einstein frames, which are conformally related. We show that the scalar field equations in the Einstein frame
are reduced to the geodesics equations on the target space of the nonlinear sigma-model.
The analytical solutions in elliptical functions are obtained when the conformal couplings are given by reciprocal exponential functions.
The behavior of the scale factor in the Jordan frame is studied using numerical computations.
For certain parameters the solutions can describe an accelerated expansion. We also derive an analytical approximation in exponential functions.

\end{abstract}

\section{Introduction}
Scalar fields play an enormous role in studies of gravity, understanding dynamics of the Universe
and the physical nature of its dark sector. First, various unified models of field theories predict
the existence of scalar partners to the tensor gravity of General Relativity.
The simplest generalizations of the Einstein's theory of gravity,
in which in addition to the metric the gravitation interaction is mediated by a scalar field, are those of scalar-tensor theories.
Second, recent observational evidence \cite{DNS1}-\cite{PA} indicates that the Universe is presently
dominated by a component dubbed dark energy.
One of the approaches to account dark energy is to introduce the cosmological constant in the framework of general relativity.
However a huge and still unexplained fine-tuning  of the cosmological constant value \cite{SW} has not been understood yet.
Another widespread interpretation of dark energy is that of quintessence,
which is described by a scalar field minimally coupled to Einstein gravity rolling down some self-interaction potential \cite{RP,CDS}.
To take into account the region where the equation of state is less than $\omega = -1$,
the model with a phantom scalar field (i.e. with a negative kinetic energy), an extension of the quintessence model,  was suggested in \cite{CAL}.
A variety of works in scalar-tensor gravity are devoted to a search of an alternate explanation of dark energy \cite{CFPS}.
An additional interest in scalar-tensor theories arises from various inflationary scenarios of the early universe \cite{LS}-\cite{VF}.
Recently gravitational models with the Higgs-potential attached much attention \cite{BMSS}-\cite{BKS}.

In papers \cite{AWE1,AWE2} the AWE Hypothesis within the framework of the generalized Brans-Dicke theory with a non-universal coupling was proposed.
The original motivation for studying this type of models is related to a unified description
 of dark matter (DM) and dark energy (DE)  based on a relaxation of the weak equivalence principle on large-scales \cite{BD}-\cite{SA}.

The model contains three different sectors: gravitation, described by the metric and the fundamental Brans-Dicke field,
the visible matter (baryons, photons, etc.)  and the invisible sector, constituted by an abnormally weighting energy (AWE).
The AWE hypothesis assumes that the invisible sector experiences the background spacetime with a different gravitational strength than the ordinary matter,
which is formulated in terms  of the non-universality of the couplings to gravity for the visible and invisible sectors.
The idea of a violation of the equivalence principle for the particular case of DM appeared prior
to the numerous evidence for cosmic acceleration and the advent of DE.
Several models based on microphysics have been considered to achieve such a mass-variation for DM in particular \cite{FR}-\cite{HSOW}.

As it is known one can describe the  matter content with either the fluid or scalar field approaches.
In \cite{AWE2} the cosmological evolution was studied in a flat FLRW background using a fluid description for the matter and the AWE sectors.
It is shown that the late-time accelerated expansion may take place in the Jordan frame
as well as there is an opportunity for building an inflation mechanism.

In this paper we continue our investigations of the AWE model and aim to obtain explicit solutions.
As distinct from previous works \cite{AWE1,AWE2} assuming exponential couplings (mutually inverse) to gravity,
we describe the matter and the invisible sector  by scalar fields, which can be both ordinary or phantom ones.
The complexity having scalar fields makes difficulties for finding exact solutions.
Nevertheless, in the Einstein frame it can be shown that under the cosmological ansatz for required solutions the gravitational equations are trivial
and scalar fields equations correspond to geodesic equations on the target space of a nonlinear sigma-model \cite{GK, IMC, GI, CHE, vHK, BM}.
We show that using the sigma-model approach yields an effective one-component Lagrangian with a potential.
The model with reciprocal exponential coupling functions  can be can be turned  to a Higgs-like one.
Scalar models with Higgs-potentials inspired by string field theories have been studied recently in \cite{ABG,APV,ALV}.
We also present exact solutions in elliptic functions for this case of the coupling functions.
In gravity theories cosmological solutions in elliptic functions have been appeared in works \cite{DGZZ}-\cite{DAW}.

The paper is arranged as follows. In the next section, we describe the model of the generalized
tensor-scalar gravity  both in the Jordan and the Einstein frames.
In Sect. 3 assuming the flat FLRW background we solve the Einstein equations and
show that the scalar field equations are equivalent to the equations of motions for a sigma-model. We also present
solutions in quadratures for scalar fields with arbitrary coupling functions.
In Sect. 4  we fix the coupling functions as reciprocal exponents and derive for various sets of parameters.
In Sect.5 using numerical computations we study the behavior of the scale factor in the Jordan frame for
certain parameters and obtain its analytical approximation in exponential functions.
The conclusions are given in Sec.6.

\section{The generalized Brans-Dicke gravity}

We start by considering the action in the Jordan frame of the generalized Brans-Dicke theory introduced in \cite{AWE1,AWE2}
\begin{eqnarray}\label{1.1}
S = \frac{c^{3}}{16 \pi \tilde{G}} \int d^{4}x \sqrt{-\tilde{g}} \left\{\Phi \tilde{R} - \frac{\omega_{BD}(M(\Phi))}{\Phi}\tilde{g}^{\mu\nu}\partial_{\mu}\Phi\partial_{\nu}\Phi \right\} + \varepsilon_{1}S_{m}[\psi_{m}, \tilde{g}_{\mu\nu}] +  \nonumber\\
\varepsilon_{2}S_{a}[\psi_{a}, M^{2}(\Phi)\tilde{g}_{\mu\nu}],
\end{eqnarray}
where $\tilde{G}$ is the "bare" gravitational constant, $\tilde{g}_{\mu \nu}$ is the Jordan-frame metric coupling universally to the  ordinary matter,
$\tilde{g}$ is the determinant of the metric $\tilde{g}_{\mu\nu}$, $\tilde{R}$ is the scalar curvature build upon $\tilde{g}_{\mu\nu}$,
$\Phi$ is a scalar degree-of-freedom, $\omega_{BD}(\Phi)$ is the Brans-Dicke coupling function while $\psi_{m,a}$
are the fundamental fields entering the physical description of the matter and
abnormally weighting sectors, respectively, $\varepsilon_{i} = \pm 1$ denotes the sign of the kinetic term for the scalar fields: $\varepsilon_{i} = +1$ corresponds to a
usual scalar field with positive kinetic energy and $\varepsilon_{i} = -1$ to a phantom field, $i = 1,2$.
It should be noted that the matter action $S_{m}$ does not explicitly
depend on the scalar field $\Phi$, so the local laws of physics are those of special relativity. The presence of the non-minimal coupling $M(\Phi)$
in the sector $S_{a}$ represents a mass-variation.

To find solutions for the model (\ref{1.1}) looks to be complicated due to
the admixture of scalar and tensor degrees of freedom.
Consequently, it is convenient to rewrite the action in the so-called Einstein frame where
the tensorial $\tilde{g}_{\mu\nu}$ and scalar $\Phi$ degrees of freedom separate into a metric $g_{\mu\nu}$ and a scalar field $\varphi$.
The Jordan and the Einstein frames are related by the conformal transformation
\begin{equation}\label{1.2}
\tilde{g}_{\mu\nu} = A^{2}_{m}(\varphi)g_{\mu\nu}
\end{equation}
with the scalar field redefinition
\begin{eqnarray}\label{1.3}
3 + 2\omega_{BD} = \left(\frac{d \ln A_{m}(\varphi)}{d\varphi}\right)^{-2}, \quad M(\Phi) = \frac{A_{a}(\varphi)}{A_{m}(\varphi)}, \quad \Phi = A^{-2}_{m}(\varphi),
\end{eqnarray}
where $A_{m}(\varphi),A_{a}(\varphi)> 0$ are the non-minimal coupling functions.
Doing so, the action (\ref{1.1}) in the Einstein frame takes the form
\begin{eqnarray}\label{1.4}
S = \frac{c^{3}}{16\pi G}\int d^{4} x \sqrt{-g}\Bigl\{R[g] - 2g^{\mu\nu}\partial_{\mu}\varphi\partial_{\nu}\varphi\Bigr\}  - \int d^{4} x \sqrt{-g}\varepsilon_{1}A^{2}_{m}(\varphi)g^{\mu\nu}\partial_{\mu}\psi_{m}\partial_{\nu}\psi_{m} - \\ \nonumber
\int d^{4} x \sqrt{-g}\varepsilon_{2}A^{2}_{a}(\varphi)g^{\mu\nu}\partial_{\mu}\psi_{a}\partial_{\nu}\psi_{a},
\end{eqnarray}
where $(g_{\mu\nu})$ is the metric with the signature $(-,+,+,+)$.
 The action is similar to those of chameleon scalar fields  \cite{KW,BBDKW}
(without the self-interaction potential).

The Einstein equations for the action (\ref{1.4}) read as follows
\begin{equation}\label{1.5}
R_{\mu\nu} - \frac 12 g_{\mu\nu} R = 2 \partial_{\mu}\varphi\partial_{\nu}\varphi - g_{\mu\nu}\partial_{\alpha}\varphi\partial^{\alpha}\varphi +  \frac{8\pi G}{c^{3}}[\varepsilon_{1}T^{(m)}_{\mu\nu} + \varepsilon_{2}T^{(a)}_{\mu\nu}].
\end{equation}
The stress-energy tensors for the ordinary and abnormally weighting sectors read
\begin{eqnarray}\label{1.6}
T^{(m)}_{\mu\nu} = 2 A^{2}_{m}(\varphi) \partial_{\mu}\psi_{m}\partial_{\nu}\psi_{m} -  A^{2}_{m}(\varphi)g_{\mu\nu}\partial_{\alpha}\psi_{m}\partial^{\alpha}\psi_{m}, \\
T^{(a)}_{\mu\nu} = 2 A^{2}_{a}(\varphi)\partial_{\mu}\psi_{a}\partial_{\nu}\psi_{a} -  A^{2}_{a}(\varphi)g_{\mu\nu}\partial_{\alpha}\psi_{a}\partial^{\alpha}\psi_{a}.
\end{eqnarray}
The field equation for $\varphi$ can be written the following form
\begin{equation}\label{1.7}
\square\varphi = - \frac{4\pi G}{c^{3}}\Bigl(\varepsilon_{1}\alpha_{m}T^{(m)} + \varepsilon_{2}\alpha_{a}T^{(a)}\Bigr),
\end{equation}
with
\begin{equation}\label{1.8}
\square\varphi = \displaystyle{\frac{1}{\sqrt{-g}}\partial_{\mu}}\left(g^{\mu\nu}\sqrt{-g}\partial_{\nu}\varphi\right), \quad
T^{i} = -2A^{2}_{i}\partial_{\alpha}\psi_{i}\partial^{\alpha}\psi_{i},
\end{equation}
where $T^{i}$  is the trace of the stress-energy tensor of the sector $i = m,a$
and $\alpha_{i} = \displaystyle{\frac{d(\ln{A_{i}})}{d\varphi}}$ are
the scalar coupling strengths to the ordinary and abnormally weighting matter, respectively.

The field equations for the scalar fields $\psi_{m}$ and $\psi_{a}$ read
\begin{eqnarray}\label{1.9}
\varepsilon_{1}\displaystyle{\frac{1}{\sqrt{-g}}\partial_{\mu}}\left(A^{2}_{m}(\varphi)g^{\mu\nu}\sqrt{-g}\partial_{\nu}\psi_{m}\right) = 0,
\\ \label{1.10}
\varepsilon_{2}\displaystyle{\frac{1}{\sqrt{-g}}\partial_{\mu}}\left(A^{2}_{a}(\varphi)g^{\mu\nu}\sqrt{-g}\partial_{\nu}\psi_{a}\right) =0.
\end{eqnarray}

\section{The sigma model formalism}
Here we consider a flat Friedman-Lema\^{\i}tre-Robertson-Walker spacetime as a background
\begin{equation}\label{2.4}
ds^{2} = g_{\mu\nu}dx^{\mu}dx^{\nu} = - c^{2}dt^{2} + a^{2}(t)\delta_{ij}dx^{i}dx^{j}
\end{equation}
Owing to the presence of the coupling functions $A_{m}$ and $A_{a}$, obtaining solutions for the model  (\ref{1.4})
(especially solutions to the scalar field equations) seems  difficult.
However, under the assumption that the metric is given by (\ref{2.4}) and that
the scalar fields depend on only a single (time) coordinate, the Einstein equations (\ref{1.5}) become  trivial and  the scalar field equations (\ref{1.9})-(\ref{1.10}) reduce to the equations of motion for a geodesic curve
for the $3$-component nonlinear $\sigma$-model. To show this we rewrite the Lagrangian corresponding to the action (\ref{1.4}).
Representing the set of the scalar fields as a sigma-model source term one obtains
\begin{equation}\label{2.1}
 L = R[g] - 2h_{AB}\acute{\sigma}^{A}\acute{\sigma}^{B},
\end{equation}
where  $\sigma^{A}$ is the multiplet
\begin{eqnarray}\label{2.2}
\sigma^{A} = \left(
               \begin{array}{c}
                 \varphi \\
                  \bar{\psi}_{m} \\
                 \bar{\psi}_{a} \\
               \end{array}
             \right), \quad A = 1,2,3,  \\
\textrm{with} \quad \bar{\psi}_{m} = \frac{\sqrt{8\pi G}}{c^{3/2}}\psi_{m}, \qquad \bar{\psi}_{a} = \frac{\sqrt{8\pi G}}{c^{3/2}}\psi_{a}
\end{eqnarray}
and the matrix $(h_{AB})$, $A=1,2,3,$ reads
\begin{equation}\label{2.3}
h_{AB} = \textrm{diag}\left(1, \varepsilon_{1} A^{2}_{m}, \varepsilon_{2} A^{2}_{a}\right).
\end{equation}
Here a $\acute{}$ denotes differentiation with respect to time variable $t$.
It should be noted that the above model appeared in association with spontaneous compactification
of the extra dimensions in higher-dimensional gravity \cite{OP,GMZ}.

Owing to the homogeneity and isotropy of the FLWR-background, we have only two Einstein equations (\ref{1.5})
\begin{eqnarray}\label{2.5}
3H^{2} = h_{AB}\acute{\sigma}^{A}\acute{\sigma}^{B}, \\
\label{2.6}
2\acute{H} + 3H^{2} = - h_{AB}\acute{\sigma}^{A}\acute{\sigma}^{B},
\end{eqnarray}
where $H$ is the Hubble parameter  $H = \displaystyle{\frac{\acute{a}}{a}}$. One can immediately integrate the equations (\ref{2.5})-(\ref{2.6}) and write the result as follows
\begin{equation}\label{2.7}
a_{EF} = a_{0}[3H_{0}(t - t_{0}) + 1]^{1/3},
\end{equation}
where $a_{0}$, $H_{0}$ and $t_{0}$ are constants of integration.

Using the time variable $\tau = \displaystyle{\ln{\left(t/t_{0}\right)}}$  \cite{KA-73}, the equations of motion for the scalar fields  (\ref{1.7}), (\ref{1.9}), (\ref{1.10})
can be decoupled from the gravitational part and take the form
\begin{equation}\label{2.9a}
\frac{d(h_{AB}\dot{\sigma}^{A})}{d\tau} - \frac{1}{2}\frac{\partial h_{CB}}{\partial \sigma^{A}}\dot{\sigma}^{C}\dot{\sigma}^{B} = 0.
\end{equation}
Here and in what follows $\dot{A} = \displaystyle{\frac{dA}{d \tau}}$.
Now it is clear that equations (\ref{2.9a}) are the Lagrange equations corresponding to the following Lagrangian
\begin{equation}\label{2.13}
L_{SL} = h_{AB}\dot{\sigma}^{A}\dot{\sigma}^{B}
\end{equation}
with the energy integral of motion
\begin{equation}\label{2.14}
E_{SL} = h_{AB}\dot{\sigma}^{A}\dot{\sigma}^{B}
\end{equation}
for the nonlinear sigma-model with the metric (\ref{2.3}) and coordinates $\sigma^{A} \in \mathbb{R}^{3}$, $A = 1,2,3$, (\ref{2.2})
on the target space $\mathcal{M} = (\mathbb{R}^{3},h)$.
For the constant $h_{AB}(\varphi) = h_{AB}$ the reduction to the sigma model was proved (for a more general setup) in \cite{IMC}.
The case of diagonal $h_{AB}$ with arbitrary dependence on scalar fields in $D$  dimensions, $D\geq3$, was considered in \cite{GI}.

The variables $\sigma^{A}$, $A = 2,3$, are cyclic and the corresponding equations of motion  read
\begin{equation}\label{2.9b}
\frac{d}{d\tau}\left(h_{AB}\dot{\sigma}_{a}\right) = 0, \quad A =2,3,
\end{equation}
or in a more detailed form
\begin{eqnarray}\label{2.9}
\frac{d}{d\tau}\left(\varepsilon_{1}A^{2}_{m}\dot{\bar{\psi}}_{m}\right) = 0,
\quad
\frac{d}{d\tau}\left(\varepsilon_{2}A^{2}_{a}\dot{\bar{\psi}}_{a}\right) = 0.
\end{eqnarray}

Eqs. (\ref{2.9}) give rise the constants of motion
\begin{equation}\label{2.12}
A^{2}_{m}\dot{\bar{\psi}}_{m} = C_{m}, \qquad A^{2}_{a}\dot{\bar{\psi}}_{a} = C_{a},
\end{equation}
where $C_{m}$ and $C_{a}$ are constants of integration, which are usually interpreted as scalar charges.

Using (\ref{2.12}) the Lagrangian (\ref{2.13})  can be represented in the following form
\begin{eqnarray}\label{2.15}
L_{\varphi} = \frac{1}{2}\left(\dot{\varphi}^{2} - V({\varphi})\right),
\end{eqnarray}
where the potential is given by
\begin{equation}\label{2.16}
V(\varphi) =  \varepsilon_{1}\frac{C^{2}_{m}}{A^{2}_{m}(\varphi)} + \varepsilon_{2}\frac{C^{2}_{a}}{A^{2}_{a}(\varphi)}.
\end{equation}
The  energy integral of motion (\ref{2.14}) now looks like
\begin{equation}\label{2.17}
E_{\varphi} = \frac 12 \left(\dot{\varphi}^{2} + V({\varphi})\right)
\end{equation}
and yields the following quadrature
\begin{equation}\label{2.19}
\int^{\varphi}_{\varphi_{0}}{\frac{d \bar{\varphi}}{\sqrt{2E_{\bar{\varphi}} - V(\bar{\varphi})}}} = \tau,
\end{equation}
which defines the solutions for the scalar field $\varphi$.

Thus, we come to the effective one-component model with a massive scalar field.
In the case of arbitrary coupling functions $A_{m}$ and $A_{a}$, the exact solutions for $\varphi$, $\psi_{m}$, $\psi_{a}$
are given by quadratures (\ref{2.19}) and
\begin{equation}\label{2.20}
\psi_{m} = \frac{c^{3/2}}{\sqrt{8\pi G}}\int^{\tau}_{0} \frac{C_{m}}{A^{2}_{m}}d \bar{\tau}, \quad \psi_{a} =  \frac{c^{3/2}}{\sqrt{8\pi G}}\int^{\tau}_{0}  \frac{C_{a}}{A^{2}_{a}}d \bar{\tau}.
\end{equation}
The detailed solutions for the model (\ref{1.4}), and hence its dynamics, depend on the exact solution for the scalar field $\varphi$
which is defined by the particular form of the potential $V$ in an analogous way to the constitutive coupling function $\mathcal{A}(\varphi)$ in \cite{AWE2}.

\textbf{One-loop corrections and the sigma-model.}

Let us now specify the coupling functions
\begin{equation}\label{4.1}
A_{m} = A^{-1}_{a} = e^{k_{m}\varphi},
\end{equation}
where $k_{m}$ is the coupling strength constant to the gravitational scalar $\varphi$.

The Lagrangian of the scalar sigma-model (\ref{2.13}) has the following form
\begin{equation}\label{4.2a}
L = \dot{\varphi}^{2} + \varepsilon_{1}e^{k_{m}\varphi}\dot{\bar{\psi}}^{2}_{m} +  \varepsilon_{2}e^{k_{a}\varphi}\dot{\bar{\psi}}^{2}_{a},
\end{equation}
where $k_{m}$, $k_{a}$ are couplings related by $k_{m} = - k_{a}$.

For (\ref{4.1})  one can estimate the influence of the quantum corrections on the hierarchy between the coupling strengths. The full analysis requires considering  perturbations produced by both metric and scalar field parts (see, for example, \cite{BKRS}) and will be given in our forthcoming paper \cite{AAG}. Here, for simplicity, we confine ourselves to the discussion of perturbative expansions of the sigma-model fields following the geometric background field method based on \cite{F,HPS}. In order take into account quantum corrections, counterterms should be build from products of the Riemann tensor $R_{ABCD}$ including contractions of it such as the Ricci tensor $R_{AB}$ and the scalar  curvature $R$.

Thus, at first loop, one obtains the following redefenition of the sigma-model metric
\begin{equation}\label{4.2b}
h_{AB} \rightarrow h_{AB} + c_{1}R_{AB} + c_{2}Rh_{AB}.
\end{equation}
Using the relations for $R_{AB}$ and $R$ from (\ref{A9}) and (\ref{A12}), we can conclude that for the exponential coupling functions (\ref{4.1})  the hierarchy  between the strength of the gravitational couplings to the visible and dark sectors is protected from quantum corrections at first loop.

\section{Solutions in elliptic functions}
Here we focus our attention on the exact solutions for the couplings given by (\ref{4.1}).

The metric $h$ given by Eq.(\ref{2.3}) defined on the target space $\mathcal{M}$ can be written as follows
\begin{equation}\label{4.2}
h = d\varphi \otimes d \varphi + \varepsilon_{1}e^{2k_{m}\varphi}d \bar{\psi}_{m}\otimes d\bar{\psi}_{m} + \varepsilon_{2}e^{-2k_{m}\varphi} d\bar{\psi}_{a} \otimes d \bar{\psi}_{a}.
\end{equation}

The form of coupling functions (\ref{4.1}) is motivated by two features.  First, in \cite{VDHS} it was proved that the target space $\mathcal{M}= (\mathbb{R}^{3},h)$
with the metric (\ref{4.2}) is a homogeneous space isomorphic to the coset space $G/H$, where $G$ is the isometry group of $\mathcal{M}$ and $H$ is the isotropy subgroup of $G$. Thus, in this case one can find solutions to  the geodesic equations (\ref{2.9a}). Second, it was shown in \cite{AWE2} that  cosmic acceleration in the Jordan frame requires an inverse proportionality of $A_{m}$ and $A_{a}$.
It should be noted that the exponential coupling functions give us a target space with constant curvature $R = 2k^{2}_{m}$(see, Appendix A).

The quadrature (\ref{2.19}) takes the form now
\begin{equation}\label{4.3}
\int^{\varphi}_{\varphi_{0}}{\frac{d \bar{\varphi}}{\sqrt{2E_{\bar{\varphi}} - \varepsilon_{1}C^{2}_{m}e^{-2k_{m}\bar{\varphi}} - \varepsilon_{2}C^{2}_{a}e^{2k_{m}\bar{\varphi}}}}} = \tau.
\end{equation}
It is worth noting that a replacement $\varphi =  \varphi + \varphi_{0}$, where $\varphi_{0}$ is a certain constant, yields to the $\sinh$-Gordon equation, which is well known in quantum field theory \cite{FMS} and presents the simplest integrable  model of the affine Toda field theory, based on the root data of the Lie algebra $a^{(1)}_{1}$ \cite{MOP}.

Introducing a new variable $z$ and redefining the parameters
\begin{equation}\label{4.4}
z = e^{k_{m}\varphi}, \quad a = - \varepsilon_{2}k^{2}_{m}C^{2}_{a},\quad b = 2k^{2}_{m}E_{\varphi},\quad c = - \varepsilon_{1}k^{2}_{m}C^{2}_{m},
\end{equation}
one can rewrite  (\ref{4.3}) in the following form
\begin{equation}\label{4.5}
\int^{z}_{z_{0}}{\frac{d\bar{z}}{\sqrt{a\bar{z}^{4} + b\bar{z}^{2} + c}}} = \tau.
\end{equation}

We can easily recognize  in  Eq.(\ref{4.5}) the equation of motion of the Higgs scalar field  considered as the inflaton \cite{BS},\cite{BKS}.

Thus, the case of exponential coupling functions  (\ref{4.1}) gives rise to a quartic polynomial for the integrand  (\ref{4.5}) and
the solution for $\varphi$ can be obtained in the terms of elliptic functions \cite{AS, NIA}.
Depending on sets of the parameters $a$, $b$ and $c$, the roots of the polynomial define the following five cases of  solutions.

\textbf{(i)}

When the parameters obey the restrictions
\begin{equation} \label{4.8}
 a< 0, \quad c < 0,\quad b > 0,\quad b^{2} >4ac
\end{equation}
the roots of the polynomial $az^{4} + bz^{2} + c$ are given by
  \begin{equation}\label{4.6}
 \rho^{2} = \frac{b - \sqrt{b^{2} - 4ac}}{2|a|}>0, \quad \lambda^{2} = \frac{b + \sqrt{b^{2} - 4ac}}{2|a|}>0
   \end{equation}
where
\begin{equation}\label{4.7}
\rho^{2} < \lambda^{2},\quad  0 < \rho \leq z \leq \lambda.
\end{equation}

Then Eq.(\ref{4.5}) can be rewritten in the form

\begin{equation}\label{4.9}
  \int^{z}_{z_{0}}{\frac{d\bar{z}}{\sqrt{(\bar{z}^{2}-\rho^{2})(\lambda^{2} - \bar{z}^{2})}}} = \sqrt{|a|}\tau,
\end{equation}

The latter equation  can be brought to the form
\begin{eqnarray}\label{4.10}
\frac{1}{\lambda}F\left(\arcsin \left[\frac{\lambda}{z}\sqrt{\frac{z^{2} - \rho^{2}}{\lambda^{2} - \rho^{2}}}\right],\frac{\sqrt{\lambda^{2} - \rho^{2}}}{\lambda}\right)  =  \sqrt{|a|}\tau,
\end{eqnarray}
 where $F(u , k)$ is an elliptical integral of the first kind with argument $u$ and modulus $k$; $\tau_{0}$ is the constant of integration.

The conditions (\ref{4.8}) correspond to the scalar fields $\psi_{m}$ and $\psi_{a}$ with ordinary kinetic terms  ($\varepsilon_{i} = +1$, $i = 1,2$) and 
 positive energy $E_{\varphi}$. In order to write the solution for the scalar field $\varphi$, one needs to find the inverse function to the elliptic integral in (\ref{4.10}),  i.e. the Jacobi elliptic function.  The solution to the scalar field $\varphi$ is

\begin{eqnarray}\label{4.12}
\varphi = \displaystyle{ \frac{1}{k_{m}}\ln\left[\frac{\lambda\rho}{\sqrt{\lambda^{2}  - \lambda^{2}\textrm{sn}^{2}\left[\sqrt{|a|}\lambda\tau,k \right] + \rho^{2}\textrm{sn}^{2}\left[\sqrt{|a|}\lambda\tau,k\right]}}\right]},
\end{eqnarray}
where $\textrm{sn}\left[\sqrt{|a|}\lambda \tau, k\right]$ is the elliptic sine function with  modulus $k =\displaystyle{ \frac{\sqrt{\lambda^{2} - \rho^{2}}}{\lambda}}$.

The coupling functions can be presented  now as follows
\begin{eqnarray}\label{4.13}
A_{m}(\varphi) = A^{-1}_{a}(\varphi) = \displaystyle {\frac{\lambda\rho}{\sqrt{\lambda^{2}  - \lambda^{2}\textrm{sn}^{2}\left[\sqrt{|a|}\lambda\tau,k \right] + \rho^{2}\textrm{sn}^{2}\left[\sqrt{|a|}\lambda\tau,k\right]}}}.
\end{eqnarray}

\textbf{(ii)}

In this case we consider the parameters $a < 0$, $c > 0$ and $b$ arbitrary.
The roots of the polynomial are defined by
 \begin{equation}\label{4.15}
  \rho^{2} = \frac{b + \sqrt{b^{2} - 4ac}}{2|a|}, \quad \lambda^{2} = \frac{-b + \sqrt{b^{2} -4ac}}{2|a|}
  \end{equation}
and
 \begin{equation}\label{4.16}
0  < z \leq \rho.
\end{equation}
Eq. (\ref{4.5}) reads now
  \begin{equation}\label{4.14}
  \int^{z}_{z_{0}}{\frac{d\bar{z}}{\sqrt{(\bar{z}^{2} + \lambda^{2})(\rho^{2} - \bar{z}^{2})}}} = \sqrt{|a|}\tau
  \end{equation}
and can be rewritten in the form
\begin{eqnarray}\label{4.17}
\frac{1}{\sqrt{\lambda^{2} + \rho^{2}}}F\left(\arcsin{\left[\frac{z}{\rho}\sqrt{\frac{\lambda^{2} + \rho^{2}}{z^{2}+\lambda^{2}}}\right]}, \frac{\rho}{\sqrt{\lambda^{2} + \rho^{2}}}\right)   =  \sqrt{|a|}\tau.
\end{eqnarray}
From (\ref{4.17}) one obtains
\begin{eqnarray}\label{4.18}
\varphi =  \frac{1}{k_{m}}\ln\left[ \frac{\rho \lambda\textrm{sn}\left[\sqrt{|a|(\lambda^{2} + \rho^{2})}\tau,k \right]}{\sqrt{\lambda^{2} + \rho^{2} - \rho^{2} \textrm{sn}^{2}\left[\sqrt{|a|(\lambda^{2} + \rho^{2})}\tau, k\right]}}\right],
\end{eqnarray}

where the modulus $k = \displaystyle{\frac{\rho}{\sqrt{\lambda^{2} + \rho^{2}}}}$.

The corresponding coupling functions are given by
\begin{eqnarray}\label{4.19}
A_{m}(\varphi) = A^{-1}_{a}(\varphi) =\frac{\rho \lambda\textrm{sn}\left[\sqrt{|a|(\lambda^{2} + \rho^{2})}\tau,k \right]}{\sqrt{\lambda^{2} + \rho^{2} - \rho^{2} \textrm{sn}^{2}\left[\sqrt{|a|(\lambda^{2} + \rho^{2})}\tau, k\right]}}.
\end{eqnarray}

The choice of parameters $a$, $b$ and $c$  corresponds to a case with a phantom scalar field for the matter sector ($\varepsilon_{1} = -1$) and a scalar field
with an ordinary kinetic term for the AWE-sector ($\varepsilon_{2} = +1$). The energy of the scalar field $\varphi$ can be either positive or negative.

  \textbf{(iii)}

Here $b$ can be either positive or negative as in the previous case, while the parameters $a$ and $c$ obey
\begin{equation}\label{4.20}
a > 0, \quad c < 0.
\end{equation}
 The  roots of the integrand (\ref{4.5}) are defined by
 \begin{equation}\label{4.21}
  \rho^{2} = \frac{b + \sqrt{b^{2} - 4ac}}{2a}, \quad \lambda^{2} = \frac{-b + \sqrt{b^{2} -4ac}}{2a},
  \end{equation}
with
 \begin{equation}\label{4.22}
   0 < \lambda \leq z < \infty.
\end{equation}

Equation (\ref{4.5}) can be rewritten in the following form
  \begin{equation}\label{4.23}
  \int^{z}_{z_{0}}{\frac{d\bar{z}}{\sqrt{(\bar{z}^{2} + \rho^{2})(\bar{z}^{2} - \lambda^{2})}}} = \sqrt{a}\tau.
  \end{equation}
and represented as follows
\begin{eqnarray}\label{4.24}
\frac{1}{\sqrt{\lambda^{2} + \rho^{2}}} F\left(\arccos{\left(\frac{\lambda}{z}\right)},\frac{\rho}{\sqrt{\lambda^{2} + \rho^{2}}}\right) =
\sqrt{a}\tau.
\end{eqnarray}

Then the solution for the scalar field $\varphi$ reads
\begin{eqnarray}\label{4.25}
\varphi =  \frac{1}{k_{m}}\ln\left(\frac{\lambda}{\textrm{cn}\left(\sqrt{a(\lambda^{2} + \rho^{2})}\tau,k\right)}\right),
\end{eqnarray}
where $\textrm{cn}\left(\sqrt{a(\lambda^{2} + \rho^{2})}\tau, k\right)$ is  the Jacobi elliptic cosine function with the modulus $k = \frac{\rho}{\sqrt{\lambda^{2} + \rho^{2}}}$.

The coupling functions read
\begin{eqnarray}\label{4.26}
A_{m}(\varphi) = A^{-1}_{a}(\varphi) = \frac{\lambda}{\textrm{cn}\left(\sqrt{a(\lambda^{2} + \rho^{2})}\tau,k\right)}.
\end{eqnarray}

Solutions (\ref{4.25})-(\ref{4.26}) correspond to a model with a usual scalar field $\psi_{m}$ (due to $\varepsilon_{1}= +1$)
and a phantom one $\psi_{a}$ (due to $\varepsilon_{2} =-1$). The scalar field energy $E_{\varphi}$ can be either positive or negative.

\textbf{(iv)}

In this case, the parameters are restricted by
\begin{equation}\label{4.27}
a > 0,\quad c >0,\quad b > 0, \quad b^{2} > 4ac.
\end{equation}
The roots are then given by
\begin{equation}\label{4.28}
  \rho^{2} = \frac{b + \sqrt{b^{2} - 4ac}}{2a}, \quad \lambda^{2} = \frac{b - \sqrt{b^{2} - 4ac}}{2a},
  \end{equation}
where
\begin{equation}\label{4.29}
0 < z < \infty, \quad 0 < \lambda^{2} < \rho^{2}.
\end{equation}
Eq.(\ref{4.5}) is rewritten in the form
\begin{equation}\label{4.30}
  \int^{z}_{z_{0}}{\frac{d\bar{z}}{\sqrt{(\bar{z}^{2} +  \lambda^{2})(\bar{z}^{2} + \rho^{2})}}} = \sqrt{a}\tau
  \end{equation}
and can be represented as follows
\begin{equation}\label{4.31}
\frac{1}{\rho}F\left(\textrm{arctan}{\left(\frac{z}{\rho}\right)}, \sqrt{\frac{\rho^{2} - \lambda^{2}}{\rho}}\right) = \sqrt{a}\tau.
\end{equation}

The solution for $\varphi$ reads
\begin{eqnarray}\label{4.32}
\varphi = \frac{1}{k_{m}}\ln\left(\rho\textrm{sc}(\sqrt{a}\lambda\tau, k)\right),
        \end{eqnarray}
where $\textrm{sc}\left(\sqrt{a}\lambda \tau,k\right)$ is the Jacobi elliptic function which can be written
as the ratio of the elliptic sine function to the elliptic cosine function with modulus $k = \sqrt{\frac{\rho^{2} - \lambda^{2}}{\rho}}$.

The coupling functions are given by
\begin{eqnarray}\label{4.33}
A_{m}(\varphi) = A^{-1}_{a}(\varphi) = \rho \textrm{sc}(\sqrt{a}\lambda\tau,k).
\end{eqnarray}

Owing to (\ref{4.27}), the  matter and AWE-sectors are described by phantom fields $\psi_{m}$, $\psi_{a}$ since $\varepsilon_{1} = -1$, $\varepsilon_{2} = -1$, while  the energy of the field $\varphi$ is positive ($E_{\varphi}>0$).

\textbf{(v)}

In this case, the parameters obey
\begin{equation}\label{4.34}
a > 0,\quad c > 0, \quad b < 0,\quad b^{2}> 4ac.
\end{equation}

The integrand roots are defined by
  \begin{equation}\label{4.35}
  \rho^{2} = \frac{-b - \sqrt{b^{2} - 4ac}}{2a} > 0, \quad \lambda^{2} = \frac{ - b + \sqrt{b^{2} -4ac}}{2a} >0,
  \end{equation}
with
\begin{equation}\label{4.37}
0  <  \rho < \lambda < z.
\end{equation}
Equation (\ref{4.5}) can be represented as follows
  \begin{equation}\label{4.36}
  \int^{z}_{z_{0}}{\frac{d\bar{z}}{\sqrt{(\bar{z}^{2} - \rho^{2})(\bar{z}^{2} -  \lambda^{2})}}} = \sqrt{a}\tau.
  \end{equation}

Using elliptic integrals of the first kind one arrives at
\begin{equation}\label{4.38}
\frac{1}{\lambda}F\left(\arcsin\left[\sqrt{\frac{z^{2} - \lambda^{2}}{z^{2} - \rho^{2}}}\right],\frac{\rho}{\lambda}\right) = \sqrt{a}\tau.
\end{equation}

The solution for the scalar field $\varphi$ is given by
\begin{eqnarray}\label{4.39}
\varphi = \frac{1}{k_{m}}\ln\left(\sqrt{\frac{\rho^{2}\textrm{sn}^{2}(\sqrt{a}\lambda \tau, k) - \lambda^{2}}{ \textrm{sn}^{2}\left(\sqrt{a}\lambda\tau,k\right) - 1}} \right).
\end{eqnarray}

The corresponding coupling functions are
\begin{eqnarray}\label{4.40}
A_{m}(\varphi) = A^{-1}_{a}(\varphi)= \sqrt{\frac{\rho^{2}\textrm{sn}^{2}(\sqrt{|a|}\lambda \tau, k) - \lambda^{2}}{ \textrm{sn}^{2}\left(\sqrt{a}\lambda\tau,k\right) - 1}}.
\end{eqnarray}

As in the previous case, both $\psi_{m}$ and $\psi_{a}$ are phantom fields ($\varepsilon_{1} = -1$, $\varepsilon_{2} = -1$),
but  the energy of the gravitational scalar is now negative ($E_{\varphi}<0$).

Thus, all generic solutions for the scalar fields $\varphi$ (\ref{4.12}), (\ref{4.17}), (\ref{4.25}), (\ref{4.32}), (\ref{4.39}) and the coupling functions (\ref{4.13}), (\ref{4.19}), (\ref{4.26}), (\ref{4.33}), (\ref{4.40}) are of oscillating type. The effective frequency of the oscillations is determined by the rate of growth of the argument.

Let us briefly discuss the classical stability of the obtained solutions. As mentioned above, they correspond to the Lagrangian (\ref{2.15}) which describes the motion in the one-dimension potential (\ref{2.16}). Here one can use the Lyapunov's method for stability of solutions near to a point of equilibrium \cite{LYAP}. Under this method, the points are stable in the sense of Lyapunov, if at these points the first derivatives of the potential vanish and its second derivatives are greater than 0, i.e. the potential should have a minimum at the equilibrium point. Thus,  one has stability in the sence of Lyapounov for case \textbf{(i)},  where we have both kinetic terms of the positive sign ($\varepsilon_{1} = +1$ and $\varepsilon_{2} = +1$). For cases \textbf{(ii, iii, iv,v)}, the potential  does not obey the conditions required for stability. Nevertheless, the potential's behavior can be changed by quantum corrections, which will be analyzed in detail in \cite{AAG}.

\section{Back to  Jordan frame}

We recall that observable quantities are not directly obtained in the Einstein frame since physical units are universally
scaled with $A_{m}(\varphi)$. Therefore, one has to find out the behavior of the scale factor in the Jordan frame.
Under Eq.(\ref{1.2}), the scale factors in the  Jordan and the Einstein frames are related in the following way
\begin{equation}\label{5.1}
\tilde{a}_{JF} = A_{m}a_{EF},
\end{equation}
where $\tilde{a}_{JF}$ is the scale factor in the Jordan frame. Thus, we have five cases of solutions.\footnote{Here we use the time variable $t$, which is related to $\tau$ by $\ln{(t/t_{0})} = \tau$.}

\begin{description}
  \item[(a)] Using the relation for the coupling function (\ref{4.13}) from the case \textbf{(i)} (Sec.4), one gets
\begin{eqnarray}\label{5.2}
\tilde{a}_{JF} =  \displaystyle {\frac{ a_{0}\lambda\rho [3H_{0}(t- t_{0})+1]^{1/3}}{\sqrt{\lambda^{2}  - \lambda^{2}\textrm{sn}^{2}\left[\sqrt{|a|}\lambda\ln{(t/t_{0})},k \right] + \rho^{2}\textrm{sn}^{2}\left[\sqrt{|a|}\lambda\ln{(t/t_{0})},k\right]}}}.
\end{eqnarray}
  \item[(b)]
  Owing  to (\ref{4.19}), the scale factor in the Jordan frame corresponding to the case \textbf{(ii)} (Sec.4) reads
  \begin{eqnarray}\label{5.3}
  \tilde{a}_{JF} =  \frac{ a_{0}\rho \lambda [3H_{0}(t- t_{0})+1]^{1/3}\textrm{sn}\left[\sqrt{|a|(\lambda^{2} + \rho^{2})}\ln{(t/t_{0})}, k \right]}{\sqrt{\lambda^{2} + \rho^{2} - \rho^{2} \textrm{sn}^{2}\left[\sqrt{|a|(\lambda^{2} + \rho^{2})}\ln{(t/t_{0})}, k\right]}}.
  \end{eqnarray}
  \item[(c)]
  Taking into account the relation (\ref{4.26}) one obtains for the third case (Sec.4)
  \begin{eqnarray}\label{5.4}
\tilde{a}_{JF} =  \frac{  a_{0}\lambda [3H_{0}(t- t_{0})+1]^{1/3}}{\textrm{cn}\left(\sqrt{a(\lambda^{2} + \rho^{2})}\ln{(t/t_{0})},k\right)}.
\end{eqnarray}
  \item[(d)]
 Owing to (\ref{4.33}), the scale factor corresponding to  case \textbf{(iv)} (Sec.4) can be written as follows:
  \begin{equation}\label{5.5}
\tilde{a}_{JF} = a_{0}\rho [3H_{0}(t- t_{0})+1]^{1/3}\textrm{sc}(\sqrt{a}\lambda\ln{(t/t_{0})},k).
\end{equation}
 \item[(e)]
 Finally, for the fifth case  with (\ref{4.40}) (Sec.4), we have
\begin{eqnarray}\label{5.6}
\tilde{a}_{JF} = a_{0}\rho [3H_{0}(t- t_{0})+1]^{1/3}\sqrt{\frac{\rho^{2}\textrm{sn}^{2}(\sqrt{a}\lambda \ln{(t/t_{0})}, k) - \lambda^{2}}{ \textrm{sn}^{2}\left(\sqrt{a}\lambda\ln{(t/t_{0})},k\right) - 1}}.
\end{eqnarray}
\end{description}

We note, that relations (\ref{5.2})-(\ref{5.6}) express the dependence of the scale factors on the Einstein time.
The time variable in the Jordan frame is related to the time variable in the Einstein frame as follows
\begin{equation}\label{5.7}
d\tilde{t} = A_{m}dt.
\end{equation}
To find out the dependence of the scale factor $\tilde{a}_{JF}$ on $\tilde{t}$ one has to integrate (\ref{5.7})
\begin{equation}\label{5.8}
\tilde{t} - \tilde{t}_{0} = \int^{t}_{t_0}A_{m}(t')dt' \equiv B(t)
\end{equation}
and substitute the inverse function to $B(t)$, which expresses the dependence $\tilde{t}(t)$, into (\ref{5.1}):
\begin{equation}\label{5.9}
\tilde{a}(\tilde{t})_{JF} = A_{m}\left(B^{-1}(\tilde{t})\right)a_{EF}\left(B^{-1}(\tilde{t})\right).
\end{equation}

Owing to the complexity of the couplings given as combinations of elliptic functions, it appears to be difficult to integrate the right-hand side of (\ref{5.9}).
However, in the third case (\ref{4.26}) (which, as a matter of fact, can be used to describe an accelerated expansion)
one can obtain an approximate analytical solution for the scale factor in the Jordan frame with dependence on $\tilde{t}$.

Let us represent the elliptic cosine function in terms of hyperbolic functions \cite{AS}
\begin{equation}\label{6.3c}
\textrm{cn}(u,k) \approx \frac{1}{\cosh u} - \frac{1}{4}k^{\prime2}(\sinh u \cosh u - u)\frac{\sinh u}{\cosh^{2}u},
\end{equation}
where $k^{2}+ k^{\prime2}  = 1$ and the modulus of the elliptic function  $k^{2}$ is close to unity.

Consequently, for the coupling function $A_{m}$ given by Eq.(\ref{4.26}) one obtains
\begin{equation}\label{6.3d}
A_{m} = \lambda\cosh\left(\sqrt{a(\lambda^{2} + \rho^{2})}\ln{(t/t_{0})}\right)
\end{equation}
and taking into account the expression for the modulus $k^{2} = \displaystyle{\frac{\rho^{2}}{\lambda^{2} + \rho^{2}}}$
 we have the following conditions for the parameters
\begin{equation}\label{6.2b}
b > 0, \quad c \approx 0, \quad a \gg 0 \quad \textrm{or} \quad E_{\varphi} > 0, \quad C_{m} \approx 0, \quad C_{a}\gg 1.
\end{equation}

Using (\ref{5.8}) and fixing, for simplicity, the parameters by $\displaystyle{\sqrt{a(\lambda^{2} + \rho^{2})} = 1}$ and $E_{\varphi} = 1$,
one can take the time variable in the Jordan frame to be

\begin{equation}\label{6.11a}
\tilde{t} = \frac14t^{2} + \frac12\ln{(t)} -\frac14.
\end{equation}
The time variable in the Einstein frame with  dependence on $\tilde{t}$ reads

\begin{equation}\label{6.12a}
t = \exp\left[\frac{4\tilde{t}+ \lambda - \lambda\textrm{W}\left(\exp\left[\frac{\lambda + 4\tilde{t}}{\lambda}\right]\right)}{2\lambda}\right],
\end{equation}
where $\textrm{W}(z)$ is the Lambert $\mathbf{W}$-function, which is defined by the equation
\begin{equation}\label{6.12b}
W(z)e^{W(z)} = z.
\end{equation}
It worth noting that the Lambert $\mathbf{W}$-function for exact cosmological solutions arises in non-local models of stringy origin in the work \cite{ALV}.

Finally, taking into account (\ref{5.9}) and (\ref{6.12a}) one can write the scale factor in the Jordan frame as

\begin{equation}\label{6.13a}
a_{JF}(\tilde{t}) = 3^{1/3}\lambda \cosh\left(\frac{4\tilde{t}+ \lambda - \lambda\textrm{W}\left(\exp\left[\frac{\lambda + 4\tilde{t}}{\lambda}\right]\right)}{2\lambda}\right)
\exp\left[\frac{4\tilde{t}+ \lambda - \lambda\textrm{W}\left(\exp\left[\frac{\lambda + 4\tilde{t}}{\lambda}\right]\right)}{6\lambda}\right].
\end{equation}

The Hubble parameter that corresponds  to the solutions  reads
\begin{equation}\label{6.13b}
H_{JF} = \frac{2\left(1 - 3\tanh \left[ \frac{\lambda \textrm{W}\left( \exp\left[\frac{4\tilde{t} + \lambda}{\lambda}\right]\right) - \lambda - 4\tilde{t}}{2\lambda}\right]\right)}
{3\lambda\left(1+ \textrm{W}\left(\exp\left[\frac{4\tilde{t} + \lambda}{\lambda}\right]\right)\right)}
\end{equation}

Another possibility for study of the scale factor in the Jordan frame (\ref{5.9}) is to find the scale factor numerically. Here we integrate numerically the equations (\ref{5.8})-(\ref{5.9}) when the coupling functions are given by Eq.(\ref{4.26}). The numerical results presented below have been obtained thanks to a parallel \textsc{C++11} program specially designed to explore the parameter space of the solutions described in this work. To handle large accelerations of the scale factor with both good precision and computation speed, the underlying algorithm relies on the main following steps:
\begin{itemize}
\item the input parameters $k_{m}$, $C_{m}$, $C_{m}$, $E_{\varphi}$, $\varepsilon_{1}$ and $\varepsilon_{2}$ are read, checked and used to determine the associated form of the solution in terms of elliptic functions;
\item the complete shape of the potential $V(\varphi)$ according to (\ref{2.16}) is computed using an optimized version of the algorithms described in \cite{NUMREC} to evaluate the elliptic functions;
\item the boundaries $\varphi_{min}$ and $\varphi_{max}$ between which computations can be executed without floating-point issues are determined using the shape of $V(\varphi)$
\item the integration starts from $t = t_{0}$ with an adaptive time step to ensure an accurate probing of the evolution of the potential;
\item at each step $n$, the value of $\tilde{t}_{n}$ is computed using Romberg integration \cite{ROM} and the value of the energy $E_{n}$ is estimated using a Ridders derivation \cite{RID} to give an order of magnitude of the error $\displaystyle{1-E_{n}/E_{\varphi}}$
\item the integration ends when $\varphi$ reaches the boundaries of the $\left[\varphi_{min}, \varphi_{max}\right]$ interval
\end{itemize}

Comparison of the approximate analytical solution (\ref{6.13a}) with the numerical one for the same set of parameters is presented on Fig.1. It is seen that according to the chosen ansatz for the elliptic function expansion (\ref{6.3c}), the solution (\ref{6.13a}) can be used for the estimation at small times, while the numerical result, including higher order terms of the expansion, is more accurate. Both energy conservation and the approximate analytical solution for small times scales have been used to check the validity and the numerical behaviour of the integration algorithms. As shown in Fig.1, in this case, the relative error for the energy estimated under the formula (\ref{2.15}) with potential given by (\ref{2.16}) and using relations (\ref{4.25})-(\ref{4.26}) stays below $10^{-10}$, which guarantees the correctness of the exhibited solution.

\begin{figure}[tbp]
 \includegraphics[scale=0.5]{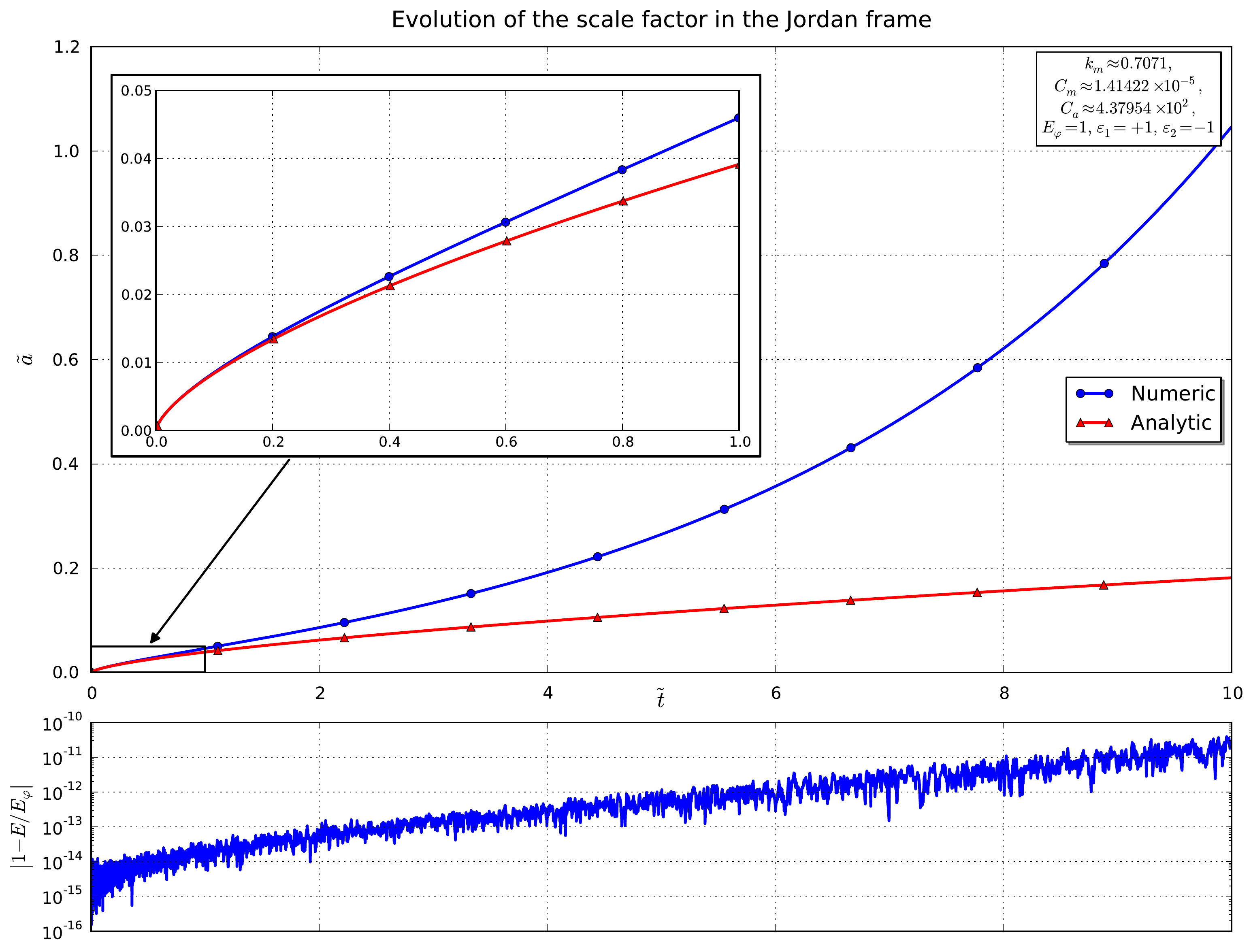}
 \caption{\textbf{Comparison of numerical and analytical solutions for the third type of solution.} In the top panel, differences between the generic computational integration and the analytical formula (\ref{6.13a}) for a set of parameters satisfying the condition (\ref{6.2b}) are presented. On short time scales, the two scale factors in the Jordan frame evolve in the same way as shown by the enlarged plot. But on larger time scales, the numerical integration exhibits an accelerated expansion, as opposed the solution in terms of Lambert $\mathbf{W}$-function. This difference comes from the fact that the analytical formula only keeps first order terms of the expansion. Nevertheless, this allows one to check the general validity of the numerical approach. The bottom plot displays another approach to control this validity by evaluating the relative error in  energy conservation at each time step. As shown here, this error stays below $10^{-10}$, thus providing a guarantee for the numerical result.}
\end{figure}

\begin{figure}[tbp]\centering{
\includegraphics[scale=0.45]{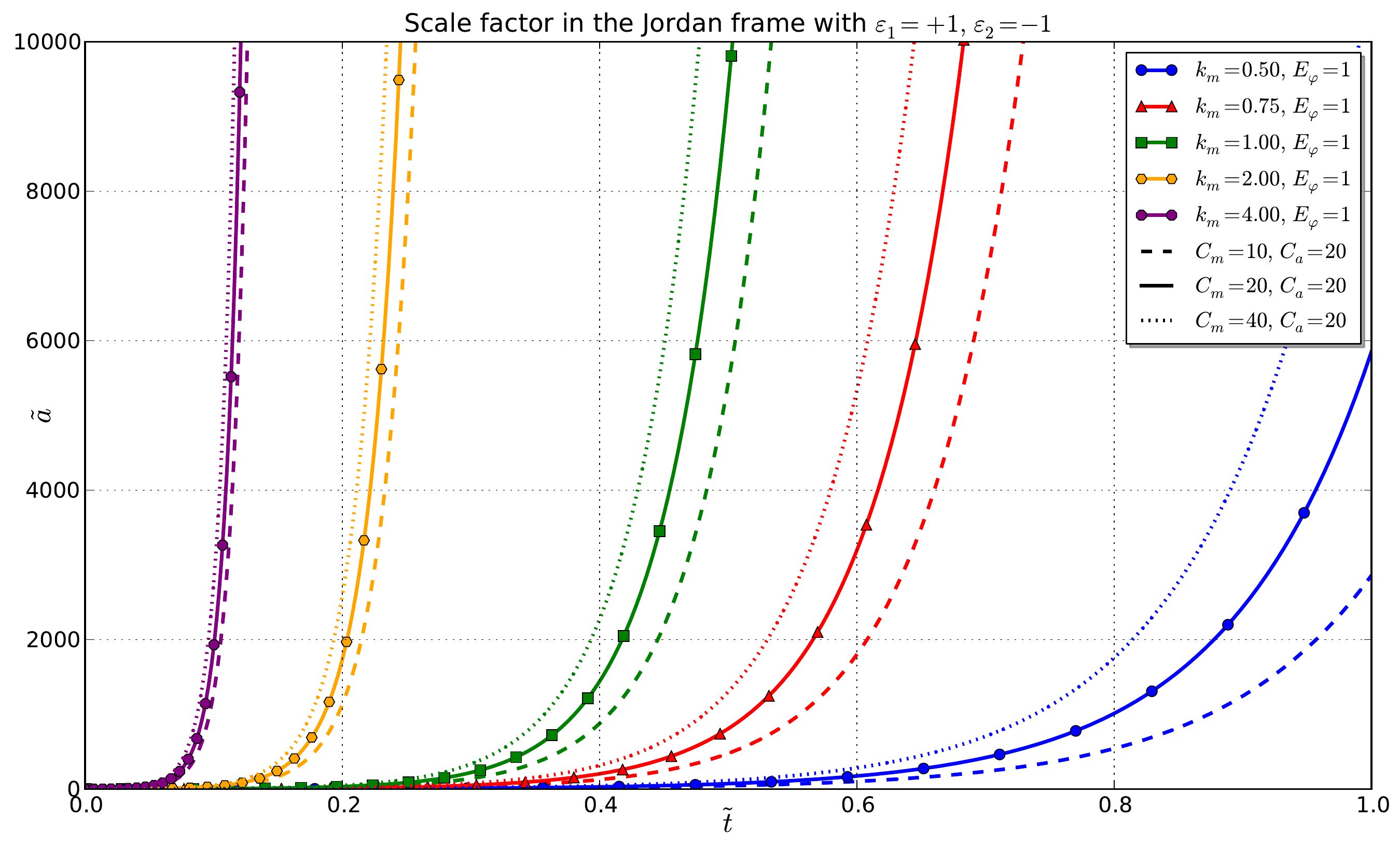}}
  \caption{\textbf{Evolution of the scale factor in the Jordan frame for several groups of parameters for the third type of solution.} Five groups of solutions are shown for several values of the coupling strength constant $k_{m}$: as expected higher values of $k_{m}$ result in faster evolution of the scale factor. For each group, three values of the scalar charges $(C_{m}, C_{a})$ are shown: for a given $k_{m}$, higher values of $C_{m}$ produce faster evolutions.
}
\end{figure}
The numerical integration allows one to move away from the condition (\ref{6.2b}); thus one can obtain larger values of the scale factor on shorter time scales. 
Fig.2 and Fig.3 illustrate the behavior of the numerical solution for the scale factor in the Jordan frame and its acceleration for other sets of parameters in the third case ($\varepsilon_{1}=+1$, $\varepsilon_{2}=-1$). It worth emphasizing that the global shape is the same as in Fig 1., but the figure shows that very large accelerations can be achieved with this model on relatively short time scales. The full  numerical analysis and computational investigations of each solution obtained in this work will be given in a forthcoming paper.

\begin{figure}[tbp]\centering{
\includegraphics[scale=0.45]{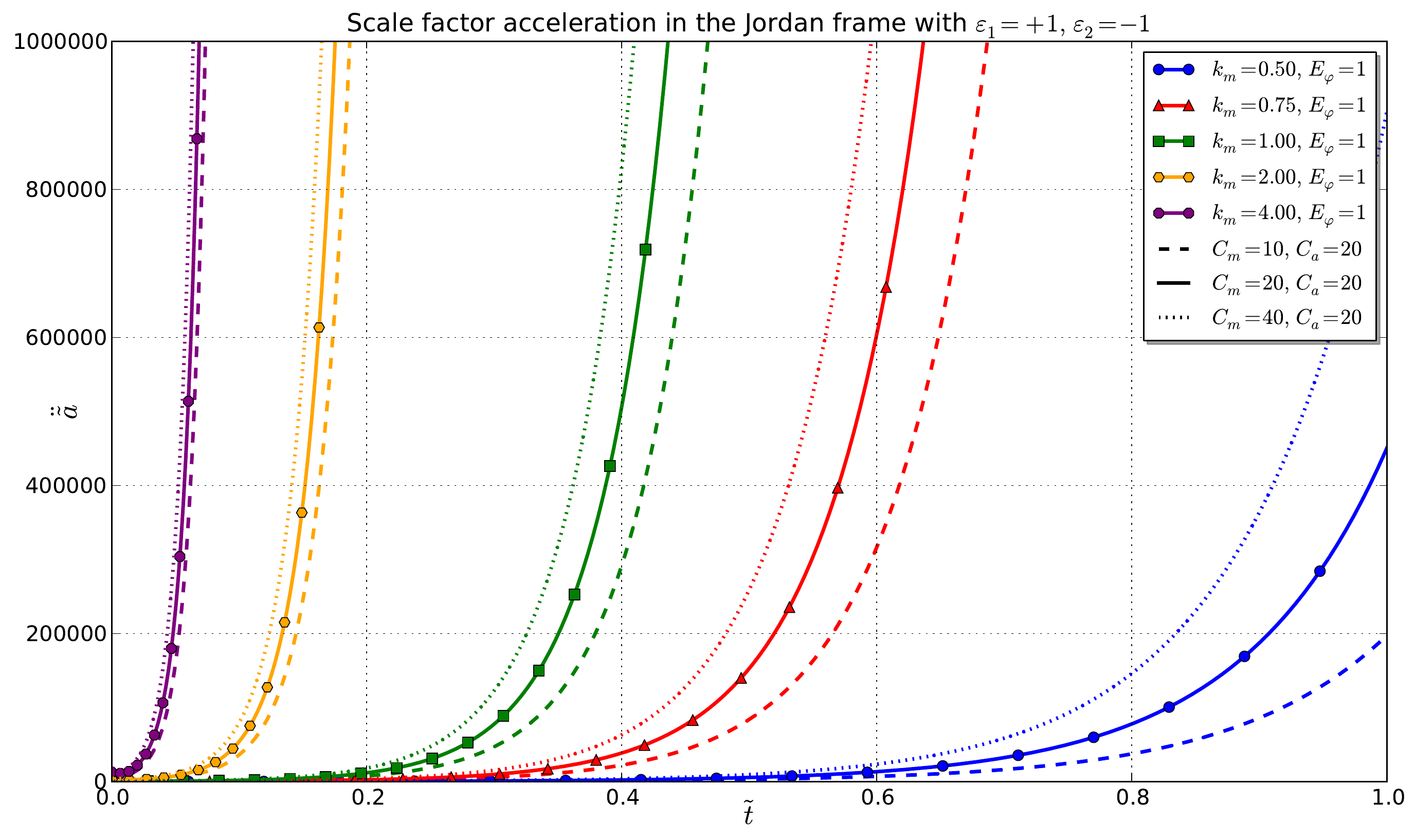}}
  \caption{\textbf{Accelerations of the scale factor in the Jordan frame for several groups of parameters for the third type of solution.} The values of the scale factor acceleration corresponding to the evolutions of Fig.2 are presented for the same five groups of parameters. Compared to the analytical formula, numerical integration allows one to probe highly accelerated solutions as in the case $k_{m} = 4$, $C_{m} = 40$, $C_{a} = 20$ and $E_{\varphi}= 1$.}
\end{figure}

\section{Conclusions}

In this work we have constructed solutions for a 4-dimensional model of the generalized Brans-Dicke theory
with a non-universal coupling,  using a combination of analytical and numerical methods.
The description of the ordinary and the AWE-sectors of the matter content are given in terms of scalar fields non-universally coupled to gravitation
via the conformal functions $A_{m}(\varphi)$ and $A_{a}(\varphi)$. The kinetic terms of these fields can have either positive or negative signs.

In the Einstein frame, we presented the considered action as a model with a sigma-model source term for the scalar fields.
Assuming the flat FLRW background, we have shown that the Einstein equations in this frame are trivial. At the same time,
 the scalar field equations correspond to geodesic equations on the target space of a sigma-model decoupled from gravitation.
 We have reduced the sigma-model to a one-component Lagrangian with a potential and have found the solutions for arbitrary coupling functions.
The solutions for the scalar fields describing the visible and invisible sectors are determined by the solution for the dilaton $\varphi$ and
the forms of the coupling functions. We have considered the case when the couplings are given by reciprocal exponential functions.
The choice of coupling functions yields a the Higgs-like equation for the dilaton.
Depending on the signs of the kinetic terms of the scalar fields, there are five cases of solutions for the scalar fields in terms of elliptic functions.
Under conformal transformations, the five cases of solutions for the dilaton yield five various forms of the scale factor in the Jordan frame.
Since the couplings are represented by combinations of elliptic functions it turns out to be difficult to derive explicit formulae for the scale factors in the Jordan frame with dependence on the Jordan time. However, we have obtained an approximate analytical solution in terms of exponential functions
when the matter sector is described by a scalar field with an ordinary kinetic term and when a phantom scalar field corresponds to the AWE-sector.
This approximate solution is defined for a special case of the parameters: a small value of the scalar charge $C_{m}$
for the scalar field describing ordinary matter and a sufficiently large value of the scalar charge $C_{a}$ related to the dark sector.
For this matter content case,  the solution in the Jordan frame has been studied numerically, see Fig.2.
By contrast to the analytical approximate solution, which is valid at very small times,
the numerical one can be suitable for the description of an accelerated expansion.

A natural extension of this work would be a detailed analysis of the conditions in which an accelerated expansion is possible.
Further study is of interest in the context of the inflationary scenario and hence would include estimation of the expansion rate of the universe and of the number of e-folds.
In addition, retracing the dynamics for the other four scale factor cases using numerical calculation is a topic of a forthcoming publication.

The present work has been focused on solutions for the flat FLWR background. At the same time,
investigation of the model for an anisotropic metric ansatz would be very attractive,
since the coupling functions which given in terms of reciprocal exponential functions might lead to interesting dynamics.
However, in the case of a non-diagonal anisotropic metric, one should use the ADM formalism  as was done in \cite{vHK}.

Here we have also  briefly discussed the influence of the quantum corrections on the hierarchy of the gravitational couplings to the matter content for the coupling functions  given by reciprocal exponential functions. In this case, we have shown that the hierarchy is protected from quantum corrections at first loop. It would be of interest to perform a detailed analysis of the higher order quantum corrections and its influence of the stability of solutions.

\section*{Acknowledgments}
We are grateful to D.S. Ageev,  A. F\"{u}zfa,  M.I. Kalinin, K.S. Stelle and S.Yu. Vernov for enlightening discussions, as well as  S.V. Bolokhov  and  V.D. Ivashchuk at the early stage of this work. AG was supported by the French Government Scholarship for the joint PhD program.

\appendix

\section{The geometric characteristics of the target manifold}
Here we present the expressions for the  Christoffel symbols, Riemann and Ricci tensors and the scalar curvature built from the metric tensor
\begin{eqnarray}\label{A1}
h = \left(
      \begin{array}{ccc}
        1 & 0 & 0 \\
        0 & A^{2}_{m}(\varphi) & 0 \\
        0 & 0 & A^{2}_{a}(\varphi) \\
      \end{array}
    \right)
\end{eqnarray}
which arises as the metric of the target space in Section 2. \\

Consider the case  $A_{m}(\varphi) = A^{-1}_{a}(\varphi)$.\\

\textbf{The nonvanishing Christoffel symbols} can be represented as

\begin{eqnarray}\label{A3}
\Gamma^{\bar{\psi}_{m}}_{\bar{\psi}_{m}\varphi} = \frac{1}{A_{m}(\varphi)}\frac{\partial A_{m}(\varphi)}{\partial \varphi}  = \alpha_{m}\qquad
\Gamma^{\varphi}_{\bar{\psi}_{m}\bar{\psi}_{m}} = -A^{2}_{m}(\varphi)\alpha_{m}, \nonumber\\
\Gamma^{\bar{\psi}_{a}}_{\bar{\psi}_{a}\varphi} = \frac{1}{A_{a}(\varphi)}\frac{\partial A_{a}(\varphi)}{\partial \varphi}= \alpha_{a} = - \alpha_{m} \qquad
\Gamma^{\varphi}_{\bar{\psi}_{a}\bar{\psi}_{a}} = -A^{2}_{a}(\varphi)\alpha_{a} = A^{-2}_{m}\alpha_{m},
\end{eqnarray}

where we denote by $\alpha_{m}$ and $\alpha_{a}$ logarithmic derivatives $\alpha_{m} = \displaystyle{ \frac{d \ln{A_{m}(\varphi)}}{d \varphi}}$ and
$\alpha_{a} = \displaystyle{\frac{d \ln{A_{a}(\varphi)}}{d \varphi}}$.

The choice of exponential coupling functions $A_{m} = e^{k_{m}\varphi}$ and $A_{a} = e^{k_{a}\varphi}$
gives rise to

\begin{eqnarray}\label{A3}
\Gamma^{\bar{\psi}_{m}}_{\bar{\psi}_{m}\varphi} = k_{m}\qquad
\Gamma^{\varphi}_{\bar{\psi}_{m}\bar{\psi}_{m}} = - k_{m} e^{2k_{m}\varphi}, \nonumber\\
\Gamma^{\bar{\psi}_{a}}_{\bar{\psi}_{a}\varphi} = k_{a} = - k_{m} \qquad
\Gamma^{\varphi}_{\bar{\psi}_{a}\bar{\psi}_{a}} = - k_{a} e^{2k_{a}\varphi} = k_{m}e^{-2k_{m}\varphi}.
\end{eqnarray}

\textbf{The nonzero components of the Riemannian tensor} in the general case read


\begin{eqnarray}\label{A5}
R^{\bar{\psi}_{m}}_{\varphi \bar{\psi}_{m} \varphi} = - R^{\bar{\psi}_{m}}_{\varphi \varphi \bar{\psi}_{m}} = - \partial_{\varphi}\alpha_{m} - \alpha^{2}_{m} = - \frac{1}{A_{m}}\frac{\partial^{2} A_{m}}{\partial \varphi^{2}}, \nonumber \\
R^{\bar{\psi}_{a}}_{\varphi \bar{\psi}_{a} \varphi} = - R^{\bar{\psi}_{a}}_{\varphi \varphi \bar{\psi}_{a}} = - \partial_{\varphi} \alpha_{a} -\alpha^{2}_{a} = -\frac{2}{A^{2}_{m}}\left(\frac{\partial A_{m}}{\partial \varphi}\right)^{2} + \frac{1}{A_{m}}\frac{\partial^{2}A_{m}}{\partial \varphi^{2}},  \nonumber \\
R^{\varphi}_{\bar{\psi}_{m}\varphi \bar{\psi}_{m}}  = - R^{\varphi}_{\bar{\psi}_{m}\bar{\psi}_{m}\varphi} = - A_{m}\frac{\partial^{2} A_{m}}{\partial \varphi^{2}}, \nonumber\\
R^{\varphi}_{\bar{\psi}_{a}\varphi \bar{\psi}_{a}}  = - R^{\varphi}_{\bar{\psi}_{a}\bar{\psi}_{a}\varphi} =  - 2 \frac{1}{A^{4}_{m}}\left(\frac{\partial A_{m}}{\partial \varphi}\right)^{2} + \frac{1}{A^{3}_{m}}\frac{\partial^{2}A_{m}}{\partial \varphi^{2}}, \nonumber \\
R^{\bar{\psi}_{a}}_{\bar{\psi}_{m}\bar{\psi}_{a}\bar{\psi}_{m}} = \left(\frac{\partial A_{m}}{\partial \varphi}\right)^{2},\qquad R^{\bar{\psi}_{m}}_{\bar{\psi}_{a}\bar{\psi}_{m}\bar{\psi}_{a}} = \frac{1}{A^{4}_{m}}\left(\frac{\partial A_{m}}{\partial \varphi}\right)^{2}.
\end{eqnarray}
According to (\ref{A3}) for
$A_{m} = e^{k_{m}\varphi}$  $A_{a} = e^{k_{a}\varphi}$,
one obtains
\begin{eqnarray}\label{A6}
R^{\bar{\psi}_{m}}_{\varphi \bar{\psi}_{m} \varphi} = - R^{\bar{\psi}_{m}}_{\varphi \varphi \bar{\psi}_{m}} = -k^{2}_{m}, \qquad R^{\bar{\psi}_{a}}_{\varphi \bar{\psi}_{a} \varphi} = - R^{\bar{\psi}_{a}}_{\varphi \varphi \bar{\psi}_{a}} = -k^{2}_{m}, \nonumber\\
R^{\varphi}_{\bar{\psi}_{m}\varphi \bar{\psi}_{m}}  = - R^{\varphi}_{\bar{\psi}_{m}\bar{\psi}_{m}\varphi} = -k^{2}_{m}e^{2k_{m}\varphi}, \quad
R^{\varphi}_{\bar{\psi}_{a}\varphi \bar{\psi}_{a}}  = - R^{\varphi}_{\bar{\psi}_{a}\bar{\psi}_{a}\varphi} = - k^{2}_{m}e^{-2k_{m}\varphi}, \nonumber\\
R^{\bar{\psi}_{a}}_{\bar{\psi}_{m}\bar{\psi}_{a}\bar{\psi}_{m}} = k^{2}_{m}e^{2k_{m}\varphi} = k^{2}_{m}e^{2k_{m}\varphi},\qquad R^{\bar{\psi}_{m}}_{\bar{\psi}_{a}\bar{\psi}_{m}\bar{\psi}_{a}} = k^{2}e^{-2k_{m}\varphi}.
\end{eqnarray}

\textbf{The nonzero components of the Ricci tensor} are given by


\begin{eqnarray}\label{A8}
R_{\varphi \varphi} = \frac{2}{A^{2}_{m}}\left(\frac{\partial A_{m}}{\partial \varphi}\right)^{2}, \quad
R_{\bar{\psi}_{m} \bar{\psi}_{m}} = \left(\frac{\partial A_{m}}{\partial \varphi}\right)^{2} - A_{m}\frac{\partial^{2} A_{m}}{\partial \varphi^{2}}, \nonumber\\
R_{\bar{\psi}_{a} \bar{\psi}_{a}} =  -\frac{1}{A^{4}_{m}}\left(\frac{\partial A_{m}}{\partial \varphi}\right)^{2} + \frac{1}{A^{3}_{m}}\frac{\partial^{2}A_{m}}{\partial \varphi^{2}}.
\end{eqnarray}

For \textbf{the exponential coupling functions} $A_{m} = e^{k_{m}\varphi}$ and $A_{a} = e^{k_{a}\varphi}$ we have
\begin{eqnarray}\label{A9}
R_{\varphi \varphi} =  2k^{2}_{m},\quad R_{\bar{\psi}_{m} \bar{\psi}_{m}} = 0,\quad
R_{\bar{\psi}_{a}\bar{\psi}_{a}} = 0.
\end{eqnarray}

Owing to (\ref{A8}) \textbf{the scalar curvature} can be written in the following form

\begin{eqnarray}\label{A11}
R = h^{\varphi\varphi}R_{\varphi\varphi} + h^{\bar{\psi}_{m}\bar{\psi}_{m}} R_{\bar{\psi}_{m}\bar{\psi}_{m}} + h^{\bar{\psi}_{a}\bar{\psi}_{a}}R_{\bar{\psi}_{a}\bar{\psi}_{a}} =  \frac{2}{A^{2}_{m}}\left(\frac{\partial A_{m}}{\partial \varphi}\right)^{2} = 2 \alpha^{2}_{m}.
\end{eqnarray}

\textbf{Remark:} It worth noting that for the case $A_{m}(\varphi)= A^{-1}_{a}(\varphi) = e^{k_{m}\varphi}$, the scalar curvature of the target space depends on the coupling strength:
\begin{equation}\label{A12}
R = 2k^{2}_{m}.
\end{equation}

\section{The direct method}
In the Einstein frame, the solutions for the scale factor and scalar fields from Sec.3 with  arbitrary coupling functions can be obtained
without resorting to the sigma-model formalism. Here we consider a direct approach for solving the field equations for the model (\ref{1.4}).
The Einstein equation in the FLRW-background can be written
\begin{equation}\label{B1}
3\frac{\dot{a}^{2}}{a^{2}}  = \dot{\varphi}^{2} + 8\pi G\left(\varepsilon_{1}A^{2}_{m}(\varphi)\dot{\psi}^{2}_{m} + \varepsilon_{2}A^{2}_{a}(\varphi)\dot{\psi}^{2}_{a}\right)
\end{equation}
\begin{equation}\label{B2}
\frac{2\ddot{a}a + \dot{a}^{2}}{a^{2}} = -\Bigl(\dot{\varphi}^{2} + 8\pi G(\varepsilon_{1}A^{2}_{m}(\varphi)\dot{\psi}^{2}_{m} + \varepsilon_{2}A^{2}_{a}(\varphi)\dot{\psi}^{2}_{a})\Bigr)
\end{equation}
The KGl dilaton equation reads now
\begin{equation}\label{B3}
\ddot{\varphi} + 3H\dot{\varphi} = 8\pi G\left(\varepsilon_{1}\alpha_{m}A^{2}_{m}(\varphi)\dot{\psi}^{2}_{m} + \varepsilon_{2}\alpha_{a}A^{2}_{a}(\varphi)\dot{\psi}^{2}_{a}\right).
\end{equation}
The KGl equation for  ordinary $\psi_{m}$ and  abnormal $\psi_{a}$ sectors can be rewritten now as
\begin{eqnarray}\label{B4}
\varepsilon_{1}\frac{d}{dt}\left(A^{2}_{m}(\varphi)a^{3}\dot{\psi}_{m}\right) = 0,
\\ \label{B5}
\varepsilon_{2}\frac{d}{dt}\left(A^{2}_{a}(\varphi)a^{3}\dot{\psi}_{a}\right) = 0.
\end{eqnarray}
The eqs. (\ref{B4})-(\ref{B5}) give rise to the constants of motion
\begin{eqnarray}\label{B6}
A^{2}_{m}(\varphi)a^{3}\dot{\psi}_{m} = c_{m}, \quad \Rightarrow \quad  \dot{\psi}_{m} = \frac{c_{m}}{A^{2}_{m}(\varphi)a^{3}},
\\ \label{B7}
A^{2}_{a}(\varphi)a^{3}\dot{\psi}_{a} = c_{a}, \quad \Rightarrow \quad  \dot{\psi}_{a} = \frac{c_{a}}{A^{2}_{a}(\varphi)a^{3}},
\end{eqnarray}
where $c_{m}$ and $c_{a}$ are some constants.
Adding Eq. (\ref{B1}) to Eq. (\ref{B2}) one obtains
\begin{equation}\label{B8}
\frac{\ddot{a}}{a} + 2\frac{\dot{a}^{2}}{a^{2}} = 0.
\end{equation}
Integration of the equation (\ref{B8}) yields the result
\begin{equation}\label{B9}
a = a_{0}(3t + C)^{1/3},
\end{equation}
where $a_{0}$ and $C$ are constants of integration.\\
As a consequence of (\ref{B6}), (\ref{B7}) and (\ref{B9}), the field equation for the scalar field $\varphi$ (\ref{B3}) now takes the form:
\begin{equation}\label{B10}
\ddot{\varphi} + \frac{3}{3t + C}\dot{\varphi} = 8\pi G\left(\varepsilon_{1}\alpha_{m}\frac{c^{2}_{m}}{A^{2}_{m}a^{6}_{0}(3t + C)^{2}} + \varepsilon_{2}\alpha_{a}\frac{c^{2}_{a}}{A^{2}_{a}a^{6}_{0}(3t + C)^{2}}\right).
\end{equation}
To solve Eq. (\ref{B10}) we should make a change of variables. So let $t + \tilde{C} = e^{u}$, where $\tilde{C} = C/3$.
Under this assumption, the equation (\ref{B10}) can be rewritten
\begin{equation}\label{B11}
\varphi^{''}_{uu} = f(\varphi),
\end{equation}
where
\begin{equation}\label{B12}
f(\varphi) = 8\pi G\left(\varepsilon_{1}\alpha_{m}\frac{c^{2}_{m}}{9A^{2}_{m}a^{6}_{0}} + \varepsilon_{2}\alpha_{a}\frac{c^{2}_{a}}{9A^{2}_{a}a^{6}_{0}}\right).
\end{equation}
Putting $\upsilon \upsilon^{'}_{\varphi} = f(\varphi)$  leads us to the following equation
\begin{equation}\label{B13}
|u| + A_{1} = \int (A_{2} + 2\int f(\varphi)d\varphi)^{-1/2} d\varphi,
\end{equation}
where $A_{1}$ and $A_{2}$ are constants.
Comparing eqs. (\ref{B13}) and (\ref{2.19}) it is easy to see the following correspondence:
\begin{equation}\label{B14}
\tau = |u|, \quad \tau_{0} = A_{1}, \quad E_{\varphi} = A_{2}, \quad V_{\varphi} = -2\int f(\varphi)d\varphi .
\end{equation}

\end{document}